\documentclass[aps,twocolumn,showpacs,preprintnumbers,amsmath,amssymb,superscriptaddress,floatfix,nofootinbib]{revtex4}

\usepackage{graphicx}
\usepackage{epsfig}
\usepackage{epstopdf}
\usepackage{subfigure}

\usepackage{amsmath}
\usepackage{amsfonts}
\usepackage{amssymb}
\usepackage{color}
\usepackage[colorlinks,
            citecolor=blue,
            anchorcolor=red,
            menucolor=red,
            linkcolor=red,
            filecolor=red,
            runcolor=red,
            urlcolor=blue,
            frenchlinks=red]{hyperref}

\begin{document}

\title{Photo-production of lowest $\Sigma^*_{1/2^-}$ state within the Regge-effective Lagrangian approach}

\author{Yun-He Lyu}
\affiliation{School of Physics and Microelectronics, Zhengzhou University, Zhengzhou, Henan 450001, China}

\author{Han Zhang}
\affiliation{School of Physics and Microelectronics, Zhengzhou University, Zhengzhou, Henan 450001, China}
	
\author{Neng-Chang Wei}
\affiliation{School of Nuclear Sciences and Technology, University of Chinese Academy of Sciences, Beijing 101408, China}

\author{Bai-Cian Ke}
\affiliation{School of Physics and Microelectronics, Zhengzhou University, Zhengzhou, Henan 450001, China}

\author{En Wang}
\affiliation{School of Physics and Microelectronics, Zhengzhou University, Zhengzhou, Henan 450001, China}

\author{Ju-Jun Xie} \affiliation{Institute of Modern Physics, Chinese Academy of
Sciences, Lanzhou 730000, China} \affiliation{School of Nuclear Sciences and Technology, University of Chinese Academy of Sciences, Beijing 101408, China} \affiliation{Southern Center for Nuclear-Science Theory (SCNT), Institute of Modern Physics, Chinese Academy of Sciences, Huizhou 516000, Guangdong Province, China}

\begin{abstract}

Since the lowest $\Sigma^{*}$ state, with quantum numbers spin-parity $J^{P} =1/2^{-}$, is far from established experimentally and theoretically, we have performed a theoretical study on the $\Sigma^*_{1/2^-}$ photo-production within the Regge-effective Lagrangian approach. Taking into account that the $\Sigma^*_{1/2^-}$ couples to the $\bar{K}N$ channel, we have considered the contributions from the $t$-channel $K$ exchange diagram. Moreover, these contributions from $t$-channel $K^*$ exchange, $s$-channel nucleon pole, $u$-channel $\Sigma$ exchange, and the contact term, are considered. The differential and total cross sections of the process $\gamma n \to K^{+}\Sigma^{*-}_{1/2^-}$ are predicted with our model parameters. The results should be helpful to search for the $\Sigma^*_{1/2^-}$ state experimentally in future.

\end{abstract}

\date{\today}

\maketitle

\section{Introduction}
\label{sec1}

The study of the low-lying excited $\Lambda^*$ and $\Sigma^*$ hyperon resonances is one of the most important issues in hadron physics. Especially, since the $\Lambda(1405)$ was discovered experimentally~\cite{Dalitz:1959dn,Alston:1961zzd}, its nature has called many attentions~\cite{Jido:2003cb,Isgur:1978xj,Capstick:1986ter,Zhang:2004xt,Callan:1985hy,Oller:2000fj}, and one explanation for $\Lambda(1405)$ is that it is a $\bar{K}N$ hadronic molecular state~\cite{Roca:2013cca,Oset:1997it,Hyodo:2011ur,Hyodo:2002pk,Guo:2012vv,Lu:2022hwm}. In addition,  the isospin $I=1$ partner of the $\Lambda(1405)$, the lowest $\Sigma^{*}_{1/2^{-}}$ is crucial to understand the light baryon spectra. At present, there is a $\Sigma^*(1620)$ with $J^P = 1/2^{-}$ listed  in the latest version of Review of Particle Physics (RPP)~\cite{PDG}. It should be stressed that the $\Sigma^*(1620)$ state is a one-star baryon resonance, and many studies indicate that the lowest $\Sigma^*_{1/2^-}$ resonance is still far from established, and its mass was predicted to lie in the range of $ 1380 \sim 1500$~MeV~\cite{Kamano:2015hxa,Zhang:2013sva,Roca:2013cca,Wu:2009nw,Wu:2009tu}. Thus, searching for the lowest $\Sigma^*_{1/2^-}$ is helpful  to understand the low-lying excited baryons with $J^P=1/2^-$ and the light flavor baryon spectra.

The analyses of the relevant data of the process $K^- p \to  \Lambda \pi^+\pi^-$  suggest that there may exist a $\Sigma^*_{1/2^{-}}$ resonance with mass about 1380~MeV~\cite{Wu:2009tu,Wu:2009nw}, which is consistent with the predictions of the unquenched quark models~\cite{Helminen:2000jb}. The analyses of the $K^{*}\Sigma$ photo-production also indicate that the $\Sigma^*_{1/2^{-}}$ is possibly buried under the $\Sigma^*(1385)$ peak with mass of 1380~MeV~\cite{Gao:2010hy}, and it is proposed to search for the $\Sigma^*_{1/2^{-}}$ in the process $\Lambda_c \to \eta\pi^+\Lambda$~\cite{Xie:2017xwx}. A more delicate analysis of the CLAS data on the process $\gamma p \to K \Sigma \pi$~\cite{CLAS:2013rjt} suggests that the $\Sigma^*_{1/2^{-}}$ peak should be around 1430~MeV~\cite{Roca:2013cca}. In Refs.~\cite{Wang:2015qta,Liu:2017hdx}, we suggest to search for such state in the processes of $\chi_{c0}(1P) \to \bar{\Sigma} \Sigma\pi$ and $\chi_{c0}(1P)\to \bar{\Lambda}\Sigma\pi$.
In addition, Ref.~\cite{Khemchandani:2018amu} has found one $\Sigma^*_{1/2^{-}}$ state with mass around 1400~MeV by solving coupled channel scattering equations, and Ref.~\cite{Kim:2021wov} suggests to search for this state in the photo-production process $\gamma p \to K^{+} \Sigma^{*0}_{1/2^-}$.   
 
 It's worth mentioning that  a $\Sigma^*(1480)$ resonance with $J^P =1/2^-$ has been listed on the previous version of RPP~\cite{ParticleDataGroup:2018ovx}. As early as 1970, the $\Sigma^*(1480)$ resonance was reported in the $\Lambda \pi^{+}$, $\Sigma \pi$, and $p\bar{K}^{0}$ channels of the $\pi^+ p$ scattering  in the Princeton-Pennsylvania Accelerator 15-in.$\sim$hydrogen bubble chamber~\cite{Pan:1969bq,Pan:1969ad}.  In 2004, a bump structure around 1480~MeV was observed in the $K^0_S p(\bar{p})$ invariant mass spectrum of the inclusive deep inelastic $ep$ scattering by the ZEUS Collaboration~\cite{ZEUS:2004lje}. Furthermore, a signal for a resonance at $1480 \pm 15$~MeV with width of $60 \pm 15$~MeV was observed in the process $pp \to K^{+}pY^{*0}$~\cite{Zychor:2005sj}. Theoretically, the $\Sigma^*(1480)$ was investigated within different models~\cite{Oh:2003fs,Oh:2007cr,Garcia-Recio:2003ejq,Oller:2006jw}. In Ref.~\cite{Oller:2006jw}, the $S$-wave meson-baryon interactions with strangeness $S=-1$ were studied within the unitary chiral approach, and one narrow pole with pole position of $1468-i \ 13 \ \rm MeV$ was found in the second Riemann sheet, which could be associated with the $\Sigma^*(1480)$ resonance. However,  the $\Sigma^*(1480)$ signals are insignificant, and  the existence of this state still needs to be confirmed within more precise experimental measurements.

As we known, the photo-production reactions have been used to study the excited hyperon states $\Sigma^*$ and $\Lambda^*$, and the Crystal Ball~\cite{CrystalBall:2001uhc,Manweiler:2008zz,Prakhov:2008dc}, LEPS~\cite{LEPS:2008azb}, and CLAS~\cite{CLAS:2013rjt} Collaborations have accumulated lots of relevant experimental data. For instance, with these data, we have analyzed the process $\gamma p\to K\Lambda^*(1405)$ to deepen the understanding of the $\Lambda^*(1405)$ nature in Ref.~\cite{Wang:2016dtb}. In order to confirm the existence of the $\Sigma^*(1480)$, we propose to investigate the process $\gamma N \to K\Sigma^*(1480)$~\footnote{Here after, we denote $\Sigma^*(1480)$ as the lowest $\Sigma^*_{1/2^-}$ state unless otherwise stated.} within the Regge-effective Lagrange approach.

Considering the $\Sigma^*(1480)$ signal was first observed in the $\pi^+\Lambda$ invariant mass distribution of the process $\pi^+ p\to \pi^+K^+\Lambda$, and the significance is about $3 \sim 4 \sigma$~\cite{Pan:1969ad}, we search for the charged $\Sigma^*(1480)$ in the process $\gamma n \to K^{+}\Sigma^{*-}_{1/2^-}$, which could also avoid the contributions of possible excited $\Lambda^*$ states. We will consider the $t$-, $s$-, $u$-channels diagrams in the Born approximation by employing the effective Lagrangian approach, and the $t$-channel $K$/$K^{*}$ exchanges terms within Regge model. Then we will calculate the differential and total cross sections of the process  $\gamma n \to K^{+}\Sigma^{*-}_{1/2^-}$ reaction, which are helpful to search for $\Sigma^*_{1/2^-}$ experimentally.

This paper is organized as follows. In Sec.~\ref{sec2a}, the theoretical formalism for studying the $\gamma n \to K^+ \Sigma^{*-}(1480)$ reactions are presented. The numerical results of total and differential cross sections and discussion are shown in Sec.~\ref{sec3}. Finally, a brief summary is given in the last section.

\section{Formalism}  \label{sec2a}

\begin{figure}[htpb] 
	\includegraphics[scale=0.6]{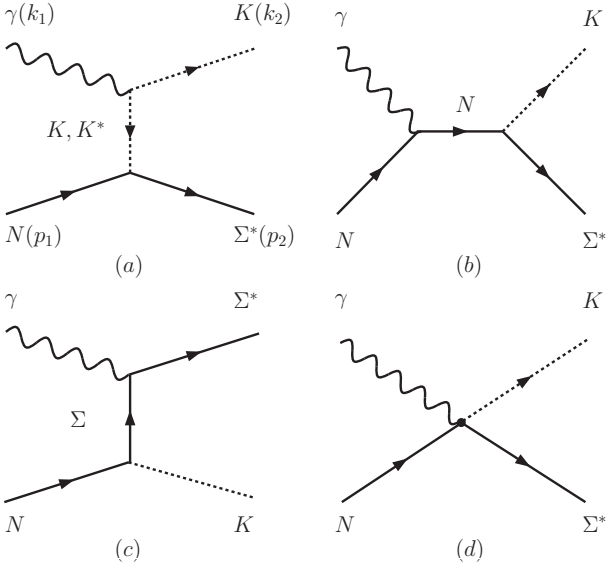}
	\vspace{0.0cm}\caption{The mechanisms of the $\gamma n \to K^+\Sigma^{*-}_{1/2^-}$ process. (a) $t$-channel $K$/$K^*$ exchange terms, (b) $s$-channel nuclear term, (c) $u$-channel $\Sigma$ exchange term, and (d) contact term. The $k_{1}$, $k_{2}$, $p_{1}$, and $p_{2}$ stand for the four-momenta of the initial photon, kaon, neutron, and $\Sigma^*(1480)$, respectively. }\label{FIG. 1. Fynmandiagrams}
\end{figure}

The reaction mechanisms of the $\Sigma^*(1480)$ $(\equiv \Sigma^*)$ photo-production process are depicted in the Fig.~\ref{FIG. 1. Fynmandiagrams}, where we have taken into account the contributions from the $t$-channel $K$ and $K^*$ exchange term, $s$-channel nucleon pole term, $u$-channel $\Sigma$ exchange term, and the contact term, respectively. 

To compute the scattering amplitudes of the Feynman diagrams
shown in Fig.~\ref{FIG. 1. Fynmandiagrams} within the effective Lagrangian approach, we use the Lagrangian densities for the electromagnetic and strong interaction vertices as used in Refs.~\cite{Wei:2022nqp,Wang:2022vln,Wang:2020mdn,Wang:2014jxb,Xie:2013mua,Kim:2021wov}
\begin{eqnarray}\label{1-7}
		\mathcal{L}_{\gamma K K} &=& -i e\left[K^{\dagger}\left(\partial_{\mu} K\right)-\left(\partial_{\mu} K^{\dagger}\right) K\right] A^{\mu},\\
	\mathcal{L}_{\gamma K K^*}&=& g_{\gamma K K^*} \epsilon^{\mu \nu \alpha \beta} \partial_\mu A_\nu\partial_\alpha K_\beta^{*}K,\\
		\mathcal{L}_{\gamma N N} &=& -e\bar{N}\left[ \gamma_{\mu}\hat{e} - \frac{\hat{\kappa}_{N}}{2M_{N}}\sigma_{\mu \nu}\partial^{\nu}\right]A^{\mu}N,\\
		\mathcal{L}_{\gamma \Sigma \Sigma^{*}} &=&\frac{e\mu_{\Sigma \Sigma^{*}}}{2M_{N}}\bar{\Sigma}\gamma_{5}\sigma_{\mu \nu}\partial^{\nu}A^{\mu}\Sigma^{*}+h.c.,\\
		\mathcal{L}_{K N\Sigma} &=& -i g_{KN\Sigma}\bar{N}\gamma_{5}\Sigma K +h.c.,\\
		\mathcal{L}_{K^{*} N\Sigma^{*}} &=& i\frac{g_{K^{*}N\Sigma^{*}}}{\sqrt{3}}\bar{K}^{*\mu}\bar{\Sigma}^{*}\gamma_{\mu}\gamma_{5}N + h.c.\\
		\mathcal{L}_{K N\Sigma^{*}} &=& g_{KN\Sigma^{*}}\bar{K}\bar{\Sigma^{*}}N+h.c.,	
\end{eqnarray}
where $e(=\sqrt{4\pi\alpha})$ is the elementary charge unit, $A^{\mu}$ is the photon filed, and $\hat{e}\equiv (1+\tau_{3})/2$ denotes the charge operator acting on the nucleon field. $\hat{\kappa}_{N} \equiv \kappa_{p} \hat{e}+\kappa_{n}(1-\hat{e})$ is the anomalous magnetic moment, and we take    $\kappa_{n}= -1.913$ for neutron~\cite{PDG}. $M_N$ and $M_{\Sigma}$ denote the masses of nucleon and the ground-state of $\Sigma$ hyperon, respectively. The strong coupling $g_{KN\Sigma}$ is taken to be 4.09 from Ref.~\cite{Rijken:2010zzb}. The $g_{\gamma K K^{*}}=0.254$~$\rm GeV^{-1}$ is determined from the experimental data of $\Gamma_{K^{*} \to K + \gamma} $~\cite{PDG} and the value of $g_{K^{*}N\Sigma^{*}}=-3.26-i0.06$ is taken from Ref~\cite{Khemchandani:2018amu}. In addition, the coupling $g_{KN\Sigma^{*}}=8.74$~GeV is taken from Ref.~\cite{Oller:2006jw}, and the transition magnetic moment $\mu_{\Sigma\Sigma^*}=1.28$ is taken from Ref.~\cite{Kim:2021wov}

With the effective interaction Lagrangian densities given above, the invariant scattering amplitudes are defined as
\begin{eqnarray}\label{8}
\mathcal{M} &=&\bar{u}_{\Sigma^{*}}(p_{2},s_{\Sigma^{*}})\mathcal{M}^{\mu}_{h}u_{N}(k_{2},s_{p})\epsilon_{\mu}(k_{1},\lambda),	
\end{eqnarray}
where $u_{\Sigma^{*}}$ and $u_{N}$ stand for the Dirac spinors, respectively, while $\epsilon_{\mu}(k_{1},\lambda)$ is the photon polarization vector and the sub-indice $h$ corresponds to different diagrams of Fig.~\ref{FIG. 1. Fynmandiagrams}.
The reduced amplitudes $\mathcal{M}^{\mu}_{h}$ are written as 
\begin{eqnarray}\label{8-14}
	\mathcal{M}^{\mu}_{K^{*}} &=& \frac{eg_{\gamma KK^{*}}g_{K^{*}N\Sigma^{*}}}{\sqrt{3}M_{K^{*}}(t-M^2_{K^{*}})}\epsilon^{\alpha\beta\mu\nu}k_{1\alpha}k_{2\beta}\gamma_{\nu}\gamma_{5},\\
	\mathcal{M}^{\mu}_{K^{-}} &=& -2i\frac{eg_{KN\Sigma^{*}}}{t-M^{2}_{K}}k_{2}^{\mu},\\
	\mathcal{M}^{\mu}_{\Sigma^{-}} &=& -i\frac{e\mu_{\Sigma\Sigma^{*}}g_{KN\Sigma}}{2M_{n}(u-M^2_{\Sigma^{*}})}(q\!\!\!/_{u}- M_{\Sigma})\sigma^{\mu \nu}k_{1\nu},\\
	\mathcal{M}^{\mu}_{n} &=& \frac{\kappa_{n}g_{KN\Sigma^{*}}}{2M_{n}(s-M_{n}^{2})}
	\sigma^{\mu\nu}k_{1\nu}(q\!\!\!/_{s}+M_{n}).   	
\end{eqnarray}	

In order to keep the full photoproduction amplitudes considered here gauge invariant, we adopt the amplitude of the contact term 
\begin{eqnarray}
\mathcal{M}^{\mu}_{c} &=& -ieg_{KN\Sigma^{*}}\frac{p_{2}^{\mu}}{p_{2}\cdot k_{1}},
\end{eqnarray}
for $\gamma n \to K^{+} \Sigma^{*-}_{1/2^-}$.

It is known that the Reggeon exchange mechanism plays a crucial role at high energies and forward angles~\cite{Donnachie:1987pu,Grishina:2005cy,Wang:2017hug,He:2014gga}, thus we will adopt  Regge model for modeling the $t$-channel $K$ and $K^{*}$ contributions by  replacing the usual pole-like Feynman propagator with the corresponding Regge propagators as follows,

\begin{eqnarray}\label{15-16}
	\frac{1}{t-M^{2}_{K}} &\to& \mathcal{F} ^{\rm Regge}_{K} \nonumber \\  &=&  \left( \frac{s}{s_{0}^{K}}\right ) ^{\alpha_{K}(t)} \frac{\pi \alpha^{\prime}_{K}}{{\rm sin}(\pi \alpha_{K}(t))\Gamma(1+\alpha_{K}(t))},\\
	\frac{1}{t-M^{2}_{K^{*}}} &\to&  \mathcal{F}^{\rm Regge}_{K^*}\nonumber \\  &=&\left( \frac{s}{s_{0}^{K^{*}}}\right ) ^{\alpha_{K^{*}}(t)} \frac{\pi \alpha^{\prime}_{K^{*}}}{{\rm sin}(\pi \alpha_{K^{*}}(t))\Gamma(\alpha_{K^{*}}(t))},
\end{eqnarray}
with $\alpha_{K}(t)=0.7~{\rm GeV}^{-2}\times(t-M_{K}^{2})$ and $\alpha_{K^{*}}(t)=1+0.83~{\rm Gev}^{-2}\times(t-M_{K^{*}}^{2})$ the linear Reggeon trajectory. The constants $s^{K}_{0}$ and $s^{K^{*}}_{0}$ are determined to be 3.0~GeV$^{2}$ and 1.5~GeV$^{2}$, respectively~\cite{Guidal:1997hy}. Here, the $\alpha'_K$ and $\alpha'_{K^*}$ are the Regge-slopes.

Then, the full photo-production amplitudes for $\gamma n \to K^+ \Sigma^{*-}_{1/2^-}$ reaction can be expressed as
\begin{eqnarray}\label{17-18}
\mathcal{M}^{\mu} &=& \left(\mathcal{M}_{K^{-}}^{\mu} + \mathcal{M}^{\mu}_{c}\right)\left(t-M_{K^{-}}^{2}\right) \mathcal{F} ^{\rm Regge}_{K}+ \mathcal{M}_{\Sigma^{-}}^{\mu}f_{u}
	 \nonumber \\ &+& \mathcal{M}_{K^{*}}^{\mu}\left(t-M_{K^{*}}^{2}\right) \mathcal{F}^{\rm Regge}_{K*} +\mathcal{M}_{n}^{\mu}f_{s},
\end{eqnarray}
While $\mathcal{F} ^{\rm Regge}_{K}$ and $\mathcal{F}^{\rm Regge}_{K*}$ stand for the Regge propagators. The form factors $f_{s}$ and $f_{u}$ are included to suppress the large momentum transfer of the intermediate particles and  describe their off-shell behavior, because the intermediate hadrons are not point-like particles.
For $s$-channel and $u$-channel baryon exchanges, we use the following form factors~\cite{Wei:2022nqp,Huang:2012xj}
\begin{eqnarray}
	f_{i}(q_{i}^{2}) = \left(\frac{\Lambda_{i}^{4}}{\Lambda_{i}^{4} + (q_{i}^{2}-M_{i}^{2})^{2}}\right)^{2},  i=s,u
	\label{21}
\end{eqnarray}
with $M_{i}$ and $q_{i}$ being the masses and four-momenta of the intermediate baryons, and the $\Lambda_{i}$ is the cut-off values for baryon exchange diagrams. In this work, we take $\Lambda_s=\Lambda_u=1.5$~GeV, and will discuss the results with different cut-off.

Finally, the unpolarized differential cross section in the center of mass (c.m.) frame for the $\gamma n \to K\Sigma^{*-}_{1/2^{-}}$ reaction reads
\begin{eqnarray}\label{22}
	\frac{d\sigma}{d\Omega} = \frac{M_{N}M_{\Sigma^{*}}|\vec{k}_{1}^{\rm c.m.}||\vec{p}_{1} ^{\rm c.m.}|}{8\pi^{2}(s-M_{N}^{2})^{2}}\sum_{\lambda,s_{p},s_{\Sigma^{*}}}^{}|\mathcal{M}|^{2},
\end{eqnarray}
where $s$ denotes the invariant mass square of the center of mass (c.m.) frame for $\Sigma^{*}_{1/2^{-}}$ photo-production. Here $\vec{k}_1^{\rm c.m.}$ and $\vec{p}_{1} ^{\rm c.m.}$ are the three-momenta of the  photon and $K$ meson in the c.m. frame, while $d\Omega = 2\pi d{{\rm cos\theta} _{\rm c.m.}}$, with $\theta_{\rm c.m.}$ the polar outgoing $K$ scattering angle.

\section{NUMERICAL RESULTS AND DISCUSSIONS} \label{sec3}

\begin{figure*}[htbp]
	\centering
			\includegraphics[scale=2.0]{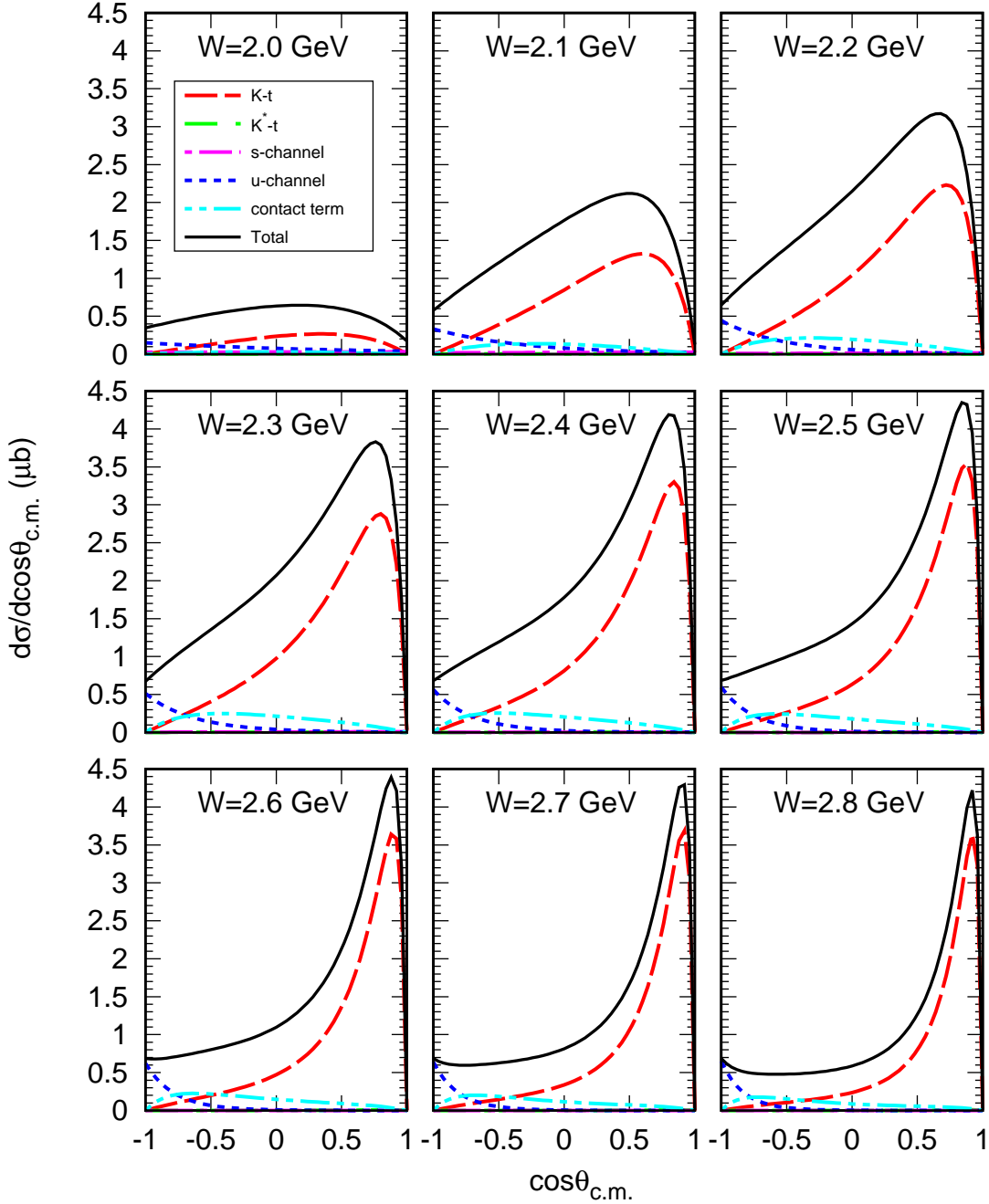}

	\caption{$\left(\rm Color \ online \right)$ $\gamma n \to K^{+}\Sigma^{*-}_{1/2^-}$ differential cross sections as a function of $\rm{cos \theta_{\rm c.m.}}$ are plotted for $\gamma n$-invariant mass intervals $\left(\rm in\ \rm GeV \ units \right)$. The black curve labeled as `Total' shows the results of all the contributions, including $t$-, $s$-, $u$- channels and contact term. The blue-dot and red-dashed curves stand for the contributions from the effective Lagrangian approach $u$- channel $\Sigma$ exchange and $t$- channel $K$ exchange mechanism, respectively. The magenta-dot-dashed and the green-dot-dashed curves show the contribution of $s$-channel and $t$-channel $K^{*}$ exchange diagrams, respectively, while the cyan-dot-dashed curve represent the contribution of the contact term. }
\label{fig:2}
\end{figure*}

In this section, we show our numerical results of the differential and total cross sections for the $\gamma n \to K^{+}\Sigma^{*-}_{1/2^-}$ reaction. The masses of the mesons and baryons are taken from RPP~\cite{PDG}, as given in Table~\ref{TAB:mass}. In addition, the mass and width of the $\Sigma(1480)$ are $M=1480\pm15$~GeV and $\Gamma=60\pm15$~GeV, respectively~\cite{ParticleDataGroup:2018ovx}.

\begin{table}[htbp]
	\begin{center}
		\caption{ \label{TAB:mass} Particle masses used in this work.  }
		\setlength{\tabcolsep}{8.5mm}{
			\begin{tabular}{cc}
				\hline\hline
				Particle                               & Mass \ $\left(\rm MeV\right)$  \\  
				\hline
				 				 $n$                     & 939.565\\		 
           $\Sigma^{-}$            & 1197.449\\
           $K^{+}$                 & 493.677   \\
				 $K^{-}$                 & 493.677   \\   
				 $K^{*}$                 & 891.66    \\
				\hline \hline
		\end{tabular}}
	\end{center}
\end{table}

\begin{figure}[htbp]
	\centering	\includegraphics[scale=0.6]{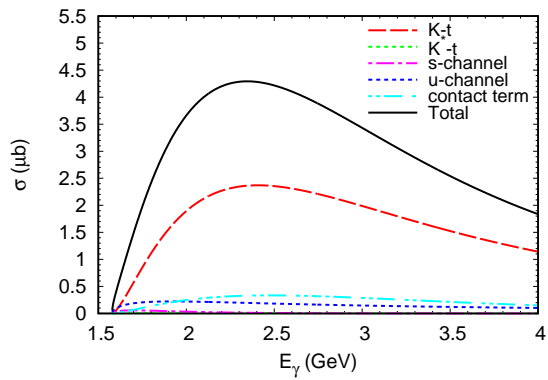}		
		\caption{ $\left(\rm Color \ online \right)$  Total cross section for $\gamma n \to K^{+}\Sigma^{*}_{1/2^-}$ is plotted as a function of the lab energy $E_{\gamma}$. The black curve labeled as `Total' shows the results of all the contributions, including $t$-,$s$-,$u$- channels and contact term. The blue-dot and red-dashed curves stand for the contributions from the effective Lagrangian approach $u$- channel $\Sigma$ exchange and $t$- channel $K$ exchange mechanism, respectively. The magenta-dot-dashed and the green-dot curves show the contribution of $s$-channel and $t$-channel $K^{*}$ exchange diagrams, respectively, while the cyan-dot-dashed curve represents the contribution of the contact term.}
	\label{fig:3}
\end{figure}

\begin{figure}[htbp]
	\centering	\includegraphics[scale=0.6]{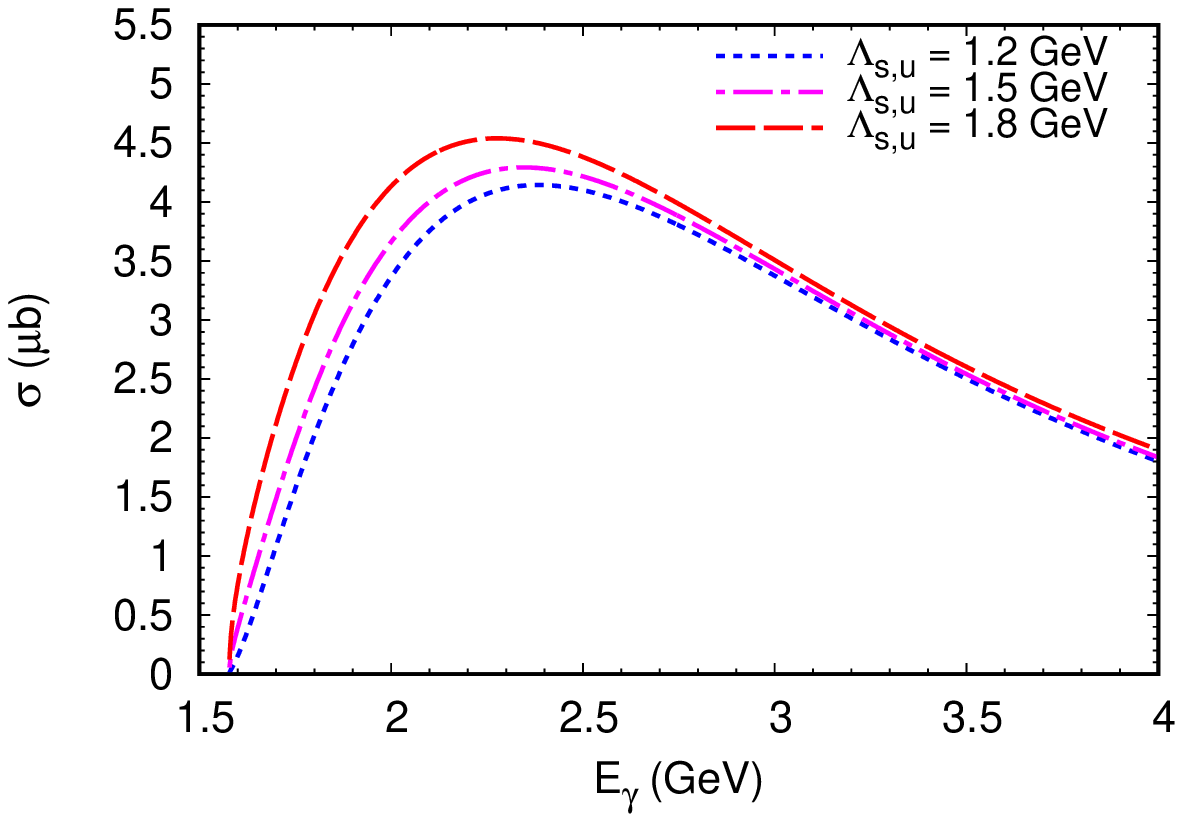}		
	
		\caption{ $\left(\rm Color \ online \right)$  Total cross section for $\gamma n \to K^{+}\Sigma^{*}_{1/2^-}$ with the cut-off $\Lambda_{s/u}=1.2$, $1.5$, and $1.8$~GeV.}
	\label{fig:dcs_cutoff}
\end{figure}
First we show the angle dependence of the differential cross sections for the $\gamma n \to K^{+}\Sigma^{*-}_{1/2^-}$ reaction in Fig.~\ref{fig:2}, where the the center-of-mass energies $W=\sqrt{s}$ varies from $2.0$ to $2.8$~GeV. The black curves labeled as `Total' show the results of all the contributions from the $t$-, $s$-, $u$-channels, and contact term. The blue-dot curves and red-dashed curves stand for the contributions from the $u$-channel $\Sigma$ exchange and $t$-channel $K$ exchange mechanism, respectively. The magenta-dot-dashed curves and the green-dot curves correspond to the contributions from the $s$-channel and $t$-channel $K^{*}$ exchange diagrams, respectively, while the cyan-dot-dashed curves represent the contribution from the contact term.  According to the differential cross sections, one can find that the $t$-channel $K$ meson exchange term plays an important role at forward angles for the process $\gamma n \to K^{+}\Sigma^{*-}_{1/2^-}$, mainly due to the Regge effects of the $t$-change $K$ exchange. 
The $K$-Reggeon
exchange shows steadily increasing behavior with ${\rm cos}\theta_{\rm c.m.}$
and falls off drastically at very forward angles. 
In addition, the $u$-channel $\Sigma$ exchange term mainly contribute to the backward angles for both processes. It should be stressed that the contribution from the $t$-channel $K^*$ exchange term is very small and could be safely neglected for the process $\gamma n \to K^{+}\Sigma^{*-}_{1/2^-}$, which is consistent with the results of Ref.~\cite{Kim:2021wov}.

In addition to the the differential cross sections, we have also calculated the total cross section of the  $\gamma n \to K^{+}\Sigma^{*-}_{1/2^-}$ reaction as a function of the initial photon energy. The results are shown in Fig.~\ref{fig:3}. The black curve labeled as `Total' shows the results of all the contributions, including $t$-, $s$-, $u$- channels and contact term. The blue-dot and red-dashed curves stand for the contributions from the $u$- channel $\Sigma$ exchange and $t$- channel $K$ exchange mechanism, respectively. The magenta-dot-dashed and the green-dot curves show the contribution of $s$-channel and $t$-channel $K^{*}$ exchange diagrams, respectively, while the cyan-dot-dashed curve represents the contribution of the contact term. For the $\gamma n \to K^{+}\Sigma^{*-}_{1/2^-}$ reaction its total cross section attains a maximum value of about $4.3 \ \mu b$ at $E_{\gamma}=2.3 \ \rm GeV$. It is expected that the $\Sigma^*(1480)$ could be observed by future experiments in the process $\gamma n \to K^{+} \Sigma^{*-}\left(1480\right) \to \Sigma^{-}\pi^{0}/\Sigma^{0}\pi^{-}/\Sigma^{-}\gamma$.

Finally, we also show the total cross section for $\gamma n \to K^{+}\Sigma^{*-}_{1/2^-}$ with the cut-off $\Lambda_{s/u}=1.2$, $1.5$, and $1.8$~GeV in Fig.~\ref{fig:dcs_cutoff}, where one can find the total cross sections are weakly dependence on the value of the cut-off. Since the precise couplings of the $\Sigma(1480)$ are still unknown, the future experiment would be helpful to constrain these couplings if the state $\Sigma(1480)$ is confirmed.
 
\section{SUMMARY} \label{sec4}

The lowest $\Sigma^{*-}_{1/2^-}$ is far from established, and its existence is important to understand the low-lying excited baryon with $J^P=1/2^-$. There are many experimental hints of the $\Sigma^*(1480)$, which has been listed in the previous version of the Review of Particle Physics. We propose to search for this state in the photoproduction process to confirm its existence.

 Assuming that the $J^{P} = 1/2^{-}$ low lying state $\Sigma^*\left(1480\right)$ has a sizeable coupling to the $\bar{K}N$ according the study of Ref.~\cite{Oller:2006jw}, we have phenomenologically investigated the $\gamma n \to K^{+} \Sigma^{*-}_{1/2^-}$ reaction by considering the contributions from the $t$-channel $K/K^*$ exchange term, $s$-channel nucleon term, $u$-channel $\Sigma$ exchange term, and contact term within the Regge-effective Lagrange approach. The differential cross sections and total cross sections for these processes are calculated with our model parameters. The total cross section of  $\gamma n \to K^{+}\Sigma^{*-}_{1/2^-}$ is about $4.3 \ \mu b$ around $E_{\gamma}=2.3 \ \rm GeV$. We encourage our  experimental colleagues to measure $\gamma n \to K^{+}\Sigma^{*-}_{1/2^-}$ process.

\section*{Acknowledgements}

This work is supported by the National Natural Science Foundation of China under Grant Nos. 12192263, 12075288, 11735003, and 11961141012, the Natural Science Foundation of Henan under Grand No. 222300420554. It is also supported by the Project of Youth Backbone Teachers of Colleges and Universities of Henan Province (2020GGJS017), the Youth Talent Support Project of Henan (2021HYTP002), the Open Project of Guangxi Key Laboratory of Nuclear Physics and Nuclear Technology, No.NLK2021-08, the Youth Innovation Promotion Association CAS.

\end{document}